\documentclass[amsmath,twocolumn,amssymb,prd,preprintnumbers,showpacs]{revtex4}
\usepackage{amsmath}
\usepackage{graphicx}
\def \be {\begin{equation}}
\def \ee {\end{equation}}
\def \bea {\begin{eqnarray}}
\def \eea {\end{eqnarray}}
\usepackage{graphicx}
\usepackage{amssymb}
\begin{document}
\title{Testing symmetries in effective models of higher derivative field theories}
\author{C. Marat Reyes}
\email[Electronic mail: ]{carlos.reyes@nucleares.unam.mx}
\affiliation{Instituto de Ciencias Nucleares,
Universidad Nacional Aut{\'o}noma de M{\'e}%
xico, 
A. Postal 70-543, 04510 M{\'e}xico D.F., M{\'e}xico}
\date{January 2009; 
published as Phys.\ Rev.\ D {\bf 80}, 105008 (2009)}
\begin{abstract}
Higher derivative field theories with interactions raise serious
doubts about their validity due to severe energy instabilities.
In many cases the implementation of a direct perturbation treatment to excise the
dangerous negative-energies from a higher
derivative field theory may lead to violations of Lorentz
and other symmetries. 
In this work we study a perturbative formulation
for higher derivative field theories that allows the construction of a low-energy
effective field theory being a genuine perturbations over the ordinary-derivative theory 
and having a positive-defined Hamiltonian.
We show that some discrete symmetries are recovered in the low-energy effective theory 
when the perturbative method to reduce the negative-energy degrees of freedom from the higher derivative theory
is applied. In particular, we focus on the higher derivative Maxwell-Chern-Simons model
which is a Lorentz invariant and parity-odd theory in $2+1$ dimensions.
The parity violation arises in the
effective action of QED$_3$ as a quantum correction from the massive fermionic sector.
We obtain the effective field theory which
remains Lorentz invariant, but parity invariant to the order considered
in the perturbative expansion.
\end{abstract}
\maketitle
\section{Introduction}
The standard model has been extremely well tested
at presently attainable energies, nevertheless it is in general regarded as an effective theory
valid up to certain high energy scale at which small imprints of a more fundamental theory
can appear. This is assumed in part because the underlying theory could
provide a solution to the ultraviolet divergences
in quantum field theories, incorporate gravity or explain other incomplete issues.
With the aim to find some of the possible low-energy effects
of the more fundamental theory, usually considered to be a quantum gravity theory or related to a unified theory,
several extended field theory models with small modifications have been constructed.
These small modifications have been proposed mainly through the use of non-higher
dimensional operators with some of them including the possibility
of Lorentz and CPT symmetry violations \cite{KOSTELECKY}.
On the contrary, models containing higher dimensional operators
have been less studied or attractive due to the many difficulties involved in their formulation \cite{PAIS-UHLENBECK};
even when they can be incorporated without any symmetry breaking.
In spite of their drawbacks presently higher derivative field theories 
continue to be strongly motivated due to the insights they are believed to provide in the 
elucidation of the more fundamental theory.
For example, the increase in the degrees of freedom 
of a self-interacting harmonic oscillator due to time derivatives of
increasing order obtained in the calculation of the effective action \cite{EFFECT-ACTION,EFFECT-ACTION2},
may serve to evidence that we are approximating a more fundamental extended object
which will become later formally defined as the wave functional in the quantum theory.

Higher derivatives were first considered in
field theories to get rid of the
infinities associated to point particles \cite{PODOLSKY}.
They can improve ultraviolet properties
in quantum field theories
\cite{THIRRING} and gravity \cite{GRAV}, although for a contrary point of view,
see \cite{RADIATIVE}. Moreover, they have been obtained from
string theory \cite{STRING}, non commutative theory
\cite{NONCOMMUTATIVE}, derivative expansions of the effective action
\cite{EFFECT-ACTION,EFFECT-ACTION2} and have been used in electrodynamics
\cite{PATH}, dark energy physics \cite{DARK,DARK2},
inflation \cite{INFLATION}, as ultraviolet regulators \cite{REG},
Lee-Wick models \cite{LEE-WICK}, and some other
contexts \cite{OTHERCONTEXTS}.    

Most of the problems with higher derivative field theories such as instability,
causality violations, nonunitary evolution and the possible emergence of
quantum states with negative-norm called ghosts states are intimately connected to the fact that
the energy has no bottom. That is, in general the higher order Hamiltonian,
the one producing temporal evolution, has
an unusual part of the spectrum taking infinite negative values in addition
to the infinite tower of positive energies.
For the noninteracting theory the negative-energy modes are decoupled from the positive
ones, they evolve independently, which eventually introduces no harm into the classical or quantum field theory.
However, when turning on the interactions, both excitations couple,
giving origin to Feynman vertices in the quantum field theory with the possibility of one particle
decaying into the other through the process of creation and annihilation.
Therefore, in compliance with energy conservation, an infinite number of positive and negative energy states
are allowed and actually favored to occur driving the system to an unlimited particle production. In consequence,
the vacuum state in the Fock representation becomes rapidly unstable.
One can try to overcome these problems by
passing to an alternative realization where the negative-energy states
are exchanged by negative-norm states, however spoiling the unitary evolution of the quantum theory.
By using path integrals techniques, it is conceivable to have positive transition amplitudes that
behave as perturbations over the ordinary ones, thereby taking  
very small departures from unitarity at low energies \cite{UNITARY}.
Also, there are proposals where these problems can be circumvented by modifying the standard internal product
in a $\mathcal{PT}$ symmetric model \cite{PT},
modifying the usual energy interpretation \cite{CHERVY}, or
using BRST symmetry \cite{RIVELLES}.
Much of the instability problem and their related issues can be understood by studying
mechanical models for which we strongly recommend the Refs. \cite{DARK2,MANNHEIM,SMILGA,SMILGAETAL}.

In many occasions \cite{EFFECTHOTD,EFFECTHOTD2,SIMON,RECENT,REYES2} one wants to consider higher derivative theories
when they describe small deviations to an ordinary-derivative theory (ordinary in the sense that they contain no more than first time-derivatives
in the Lagrangian).
In that case an additional problem may arise since, no matter how weak the higher derivative operators are coupled, in general, the
modified theory suffers an increase in the degrees of freedom \cite{OSTROGRADSKI}, 
which ultimately will depend on \emph{how much} the theory is constrained.
Therefore, a consistent perturbative formulation requires us to retain
the original number of degrees of freedom dictated by the ordinary-derivative theory and to reproduce only
the dynamical sector of the higher derivative theory that
is well defined when the higher derivative operators are taken to zero: the
perturbative sector.
In addition, in the presence of interactions these perturbative degrees of freedom should have a
stable evolution.

To illustrate some of the above problems, consider the Lagrangian
\begin{eqnarray} \label{FIRSTLAG}
\mathcal L=-\frac{1}{4}F_{\mu \nu}
F^{\mu \nu}+\frac{g}{2}\epsilon^{\alpha \beta
\gamma} (\Box A_{\alpha} )
(\partial_{\beta} A_{ \gamma}),
\end{eqnarray}
which, by using the field redefinitions
\begin{equation}
\bar A^{\mu}=\frac{1}{\sqrt 2}(A^{\mu} +gF^ {*\mu}),
\end{equation}
\begin{equation}
\widetilde A^{\mu}=\frac{g}{\sqrt 2}F^ {*\mu},
\end{equation}
can be rewritten as
\begin{eqnarray}
\mathcal L=-\frac{1}{4}\bar F_{\mu \nu}
\bar F^{\mu \nu}-\frac{1}{2}\widetilde A_{\mu}
\left(\frac{1}{g^2}+\Box \right) \widetilde A^{\mu},
\end{eqnarray}
where ${\bar F}_ {\mu \nu}=\partial_\mu \bar A_{\nu}-\partial_\nu \bar A_{\mu}$ is the strength tensor
and $F^ {*\mu}=\frac{1}{2}\epsilon^{\mu\alpha\beta} F_{\alpha \beta}$ the pseudovector dual field.
At this level we can emphasize that:

(1) The Lagrangian consists in the sum of a massless and a
massive term both associated to low-energy and high-energy degrees of freedom, respectively.
The massless Lagrangian is the only one depending linearly on $g$ so it must be parity odd.

(2) The massive field contributes with negative energy to the system
making the Hamiltonian unbounded from below.

(3) Since the low-energy and high-energy fields are decoupled from each other
we still have unitarity
and stability in the theory \cite{SMILGAETAL}.

(4) The mass of the high-energy field depends nonanalytically on the
parameter $g$, which goes to infinity when $g\to 0$.

(5) By turning on the interactions, for example, via a fourth power of the gauge field $(A_{\mu} \,A^{\mu})^2$,
the theory probably
collapses giving all the stability problems we have mentioned.

(6) In order to approach well the low-energy dynamics  
the nonperturbative massive mode has to be removed from the higher derivative theory which 
is not clear how to achieve by making field redefinitions.

In this paper we study the perturbative method for higher derivatives theories
developed in Refs. \cite{PERT-METHOD,STRING}. 
The method allows to retain the low-energy sector of the higher derivative theory 
and to construct a positive-defined effective Hamiltonian 
order by order in the expansion parameter $g$.
The perturbative method has been thoroughly applied in
\cite{PERT,METHODPERT,REYES},
and, in particular, it has been implemented as a starting point to quantize
Lorentz violating higher derivative field theories \cite{REYES}.
In this work we attempt to go a step further. 
From the observation that Lagrangian (\ref{FIRSTLAG}) describes a photon at low energies (taking the limit $g \to 0$), it sounds
plausible to recover electromagnetic symmetries 
by removing all the high-energy degrees of freedom from the higher derivative theory.
Thus, the main goal in this paper
is to test whether the symmetries of a higher derivative field theory can be modified
by applying the perturbative formulation.
It is well known that symmetries in an effective theory may depend on the scheme of approximation used to obtain the low-energy limit.
For example, a direct analytical expansion in the Hamiltonian 
or implementing a order reduction treatment using the equations of motion on the higher derivative
Lagrangian may lead to different results for Lorentz and other symmetries \cite{HAMILTONIAN,LORENTZ-INCONSISTENCIES}.
In order to implement and test the perturbative method we will consider
the Lagrangian (\ref{FIRSTLAG}).

The organization of the paper is as follows.
In Sec. II we obtain the 
solutions to the equations of motion for the higher derivative Maxwell-Chern-Simons (MCS).
We develop the Hamiltonian formulation with all the constraints included
and we exhibit the negative energies in the total energy spectrum produced by the massive field.
In Sec. III we review both the Hamiltonian formulation
for higher derivatives field theories and the perturbative method.
Section IV is the main part of this work. There by implementing the perturbative
method we derive the effective theory and we perform a complete study
of its symmetries.
In Sec. IV we give the conclusions and final comments.
For completeness we provide the reduced Hamiltonian
and the Dirac brackets in the appendix.
\section{The higher derivative gauge theory}
In this section we will consider higher derivative corrections to the Maxwell
dynamics that appear in a perturbative expansion of the effective
action in QED in $2+1$ dimensions \cite{DESER}.
We focus on the extended Chern-Simons term which is Lorentz invariant,
parity-violating, but no longer topological
as the original Chern-Simons term.
\subsection{Extended Maxwell-Chern-Simons model}
Let us consider the extended MCS Lagrangian
in $2+1$ dimensions in the Lorentz gauge
\begin{eqnarray} \label{GAUGELAGRANGIAN}
\mathcal L=-\frac{1}{4}F_{\mu \nu}
F^{\mu \nu}+\frac{g}{2}\epsilon^{\alpha \beta
\gamma} (\Box A_{\alpha} )
(\partial_{\beta} A_{ \gamma})
-\frac{1}{2}(\partial_{\mu} A^{\mu} )^2,\nonumber \\
\end{eqnarray}
where $g$ is a small coupling parameter. 

A direct calculation from the Lagrangian (\ref{GAUGELAGRANGIAN}) yields
\begin{eqnarray} \label{EXP1}
\frac{\partial  \mathcal L}
{\partial (\partial_\lambda A_{\sigma})}
=-F^{\lambda \sigma}+\frac{g}{2}\epsilon^
{\alpha \lambda \sigma}\Box A_{\alpha} -\eta^{\lambda
\sigma }(\partial_{\mu} A^{\mu} ),
\end{eqnarray}
\begin{eqnarray}
 \frac{\partial  \mathcal L}
{\partial (\partial_{\mu}\partial_\lambda A_{\sigma})}
=\frac{g}{2}\eta^{\mu \lambda}\epsilon^{\sigma
\beta \gamma}\partial_{\beta}A_{\gamma}.\label{EXP2}
\end{eqnarray}
The usual variation with respect to $A_{\sigma}$ gives the generalized Euler-Lagrange equation
\begin{eqnarray}\label{EULER-LAGRANGE}
 \frac{\partial \mathcal L}{\partial A_{\sigma}}-
 \partial_\lambda \frac{\partial  \mathcal L}
{\partial (\partial_\lambda A_{\sigma})}+\partial_
{\mu}\partial_\lambda \frac{\partial  \mathcal L}
{\partial (\partial_{\mu}\partial_\lambda A_{\sigma})}=0,
\end{eqnarray}
and substituting Eqs. (\ref{EXP1}), (\ref{EXP2}) leads to the modified Maxwell equations
\begin{eqnarray}\label{MAXWELL}
\partial_\lambda F^{\lambda \sigma}+\partial^{\sigma}
(\partial \cdot A )+
\frac{g}{2} \epsilon^{\sigma \beta \gamma} \Box
  F_{\beta \gamma}  =0,
\end{eqnarray}
which can be rewritten as
\begin{equation}\label{EQ}
 \left( \eta ^{\sigma \alpha } +g \epsilon^{\sigma \beta \alpha }
\partial_{\beta}   \right)\Box  A_{\alpha}=0.
\end{equation}
Using the plane wave ansatz $A_{\mu}(x)=\epsilon_{\mu}(k)e^{-ik \cdot x}$, we find the two solutions
\begin{eqnarray}\label{MODES}
k^2=0, \qquad    k^2=\frac{1}{g^2}.
\end{eqnarray}
Therefore, we see that the gauge field excitations are described by
a massless and a massive mode \cite{COMMENT1}. Here we will use the convention
$\eta^{\mu \nu}=\text{diag}(1,-1,-1)$
together with the notation $k^{\mu}=(k^0,{\bf k})$.
\subsection{The Hamiltonian formulation}
The Hamiltonian formulation
for systems with higher time-derivatives was developed long time ago by 
Ostrogradski \cite{OSTROGRADSKI}. Accordingly, we consider
$A_{\mu}(x)$ and
$\dot A_{\mu}(x)$ as two independent configuration field
variables with their corresponding conjugate momenta given by
\begin{eqnarray}
P^{\mu}=\frac{\partial \mathcal L}
{\partial \dot  A_{\mu}}-
 \frac{\partial \Pi^{\mu}}
{\partial t},
\qquad
\Pi^{\mu}=\frac{\partial  \mathcal L}
{\partial \ddot {A}_{\mu}},
\end{eqnarray}
which follows from Eq. (\ref{HIGHERMOMENTA}) of the next section.

By using the Eqs. (\ref{EXP1}) and (\ref{EXP2}) we can write down the conjugate momenta
\begin{eqnarray}
P^{\mu}&=&- F^{0 \mu  }-
\frac{g}{2}\epsilon^{ \mu 0 \gamma }
\Box A_{\gamma}-\eta^{0\mu}(\partial \cdot A)
-\frac{g}{2}\epsilon^{\mu \beta \gamma}
 \partial_{\beta }\dot A_{\gamma},\nonumber
\\ \Pi^{\mu}&=&\frac{g}{2} \epsilon^
{\mu \beta \gamma} \partial_{\beta }A_{\gamma},
\end{eqnarray}
which read in components
\begin{eqnarray} \label{MOM1}
P^0= -  (\partial \cdot A)
-\frac{g}{2} \epsilon^{ij} \partial_i
\dot A_j,
\end{eqnarray} 
\begin{eqnarray} \label{MOM2} 
P^i=F_{0i}+\frac{g}{2}
\epsilon^{ij} \Box A_j
 +\frac{g}{2} \epsilon^{ij} \partial_0 F_{0j},
\end{eqnarray}  
\begin{eqnarray}   \label{MOM3}
 \Pi^0=\frac{g}{2} \epsilon^{ij}
\partial_iA_j,
\end{eqnarray} 
\begin{eqnarray}  \label{MOM4}  
\Pi^i=-\frac{g}{2} \epsilon^{ij} F_{0j}.
\end{eqnarray}
We are using the conventions 
\begin{eqnarray}
\epsilon^{012}=\epsilon^{12}=+1, \qquad \epsilon_{012}=\epsilon_{12}=+1, \nonumber \\
\qquad \epsilon^{0ij}=\epsilon^{ij}, \qquad i,j=1,2.
\end{eqnarray}
Now, considering the generalized Legendre transformation $\mathcal H_C= P^ {\mu}\dot A_ {\mu}+ \Pi^{\mu} \ddot
A_{\mu}-\mathcal L$ the canonical Hamiltonian density $\mathcal H_C$ turns out to be
\begin{eqnarray} \label{HAM2}
\mathcal H_C&=&P^0 \dot A_0+ \frac{2}{g}
\epsilon^ {ij}P_i\,\Pi_j  -
P_i(\partial_i A_0 )+\frac{1}{4}F_{ij}^2
-\frac{2}{g^2}\Pi_i^2\nonumber \\
&&
-\Pi_i(\nabla^2A_i)+\frac{g}{2}\epsilon^
 {ij} \nabla^2A_0
(\partial_iA_j)+\frac{1}{2}(\partial
\cdot A )^2,\nonumber \\
\end{eqnarray}
where we have used $ P^i\dot A_i=2\epsilon^{ij}
P_i\Pi_j/g -P_i \partial_iA_0\; $ and $ F_{0i}^ 2=4\Pi^2_i/g^ 2$.

Examining Eqs. (\ref{MOM1}), (\ref{MOM3}) and (\ref{MOM4}) we deduce the four primary constraints
\begin{align}
\chi_0=\Pi^0-\frac{g}{2} \epsilon^{ij}
\partial_iA_j,& \qquad \chi_1=P^0+ (\partial \cdot A)
 +\frac{g}{2} \epsilon^{ij}
\partial_i \dot A_j,  \nonumber  \\
\varphi^i&=\Pi^i+\frac{g}{2} \epsilon^{ij}
 F_{0j}.
\end{align}
Using the canonical Poisson brackets for the extended phase space variables 
\begin{eqnarray}
\left\{ A_{\mu}(t,\mathbf{x}),P_{\nu}
(t,\mathbf{x}^{\prime})
\right\}&=&\eta
_{\mu \nu}\delta ^{3}(\mathbf{x}-
\mathbf{x}^{\prime}),\nonumber
\\
 \{ \dot A_{\mu}(t,\mathbf{x}),
 \Pi_{\nu}(t,\mathbf{x}^{\prime})
 \}&=&\eta
_{\mu \nu}\delta ^{3}(\mathbf{x}-
\mathbf{x}^{\prime}),
\end{eqnarray}
and after a straightforward calculation we find the non trivial elements of the algebra of constraints 
\begin{eqnarray}\label{CONSTRAINTALG}
\left\{  \chi_1(x),\chi_0(x^{\prime})\right\}&=&\delta ^{3}
(\mathbf{x}-\mathbf{x}^{\prime}),\nonumber
\\
 \left\{ \varphi^i(x), \chi_1(x^{\prime}) \right\}&=&-g\epsilon^{ij}
\frac{\partial }
{\partial {x}^j}\,\delta ^{3}(\mathbf{x}-\mathbf{x}^{\prime}) ,
\\
\left\{ \varphi^i(x),\varphi^j(x^{\prime})\right\}&=&g
\epsilon^{ij}\delta ^{3}(\mathbf{x}-\mathbf{x}^{\prime}).\nonumber
\end{eqnarray}
Therefore, according to Dirac classification, we see that all the primary constraints fall into
second class constraints category.

In order to search for additional constraints in the theory called secondary constraints, it is necessary to 
evolve the primary constraints with the total Hamiltonian, which
is defined adjoining the primary
constraints $\chi=\{ \chi_0,\chi_1,\varphi^i\}$ as follows
\begin{eqnarray}\label{HAMILTONIANO}
 H_T=\int d^{3}\mathbf{x}
 \left(\mathcal H_C+\sigma_0
 \chi_0+\sigma_1\chi_1
 +\alpha^i\varphi^i \right),
\end{eqnarray}
where the variables $\{\sigma_0,\sigma_1,\alpha^i\}$ play the role of Lagrange multipliers.
However, given that all primary constraints satisfy the weak condition
\begin{equation}
\left\{  \chi, H_T\right\}\approx 0,
\end{equation}
there are no generations of secondary constraints in the theory.
In addition, the Lagrange multipliers can be determined by comparing the 
Hamilton equations using the total Hamiltonian (\ref{HAMILTONIANO}) with
the original ones of Eq. (\ref{EQ}). After some algebra we find
\begin{eqnarray}
\sigma_0=\ddot A_0,
\qquad \sigma_1 =0, \qquad
\alpha^i=-\partial_0 F_{0i}.
\end{eqnarray}
In this way the total
Hamiltonian is shown to be consistent with the equations of motion (\ref{HAMILTONIANO}) and
the set of conjugate momenta (\ref{MOM1})-(\ref{MOM4}).
As an aside, let us mention that second class constraints can be imposed strongly to zero
in which case we shall require the use of
the reduced Hamiltonian $\mathcal H_R$ and the Dirac brackets given
in the appendix.
\subsection{The negative-energy mode }
To begin, let us express the general solution of Eq. (\ref{EQ}) 
as a sum of a massless and massive gauge fields as follows
\begin{equation}\label{SEPFIELD}
{A}_{\mu}(x)= {\bar A}_{\mu}(x)+{ \widetilde {B}} _{\mu}(x),
\end{equation}
such that ${\bar A}_{\mu}(x)$ and ${ \widetilde {B}} _{\mu}(x)$ 
satisfy the equations of motion
\begin{eqnarray} \label{SEPEQ1}
 \Box \, {\bar A}_{\alpha}=0, 
\end{eqnarray}
\begin{eqnarray} \label{SEPEQ2}
\left( \eta ^{\sigma \alpha } +g \epsilon^{\sigma \beta \alpha }
\partial_{\beta}   \right) {\widetilde B}_{\alpha}=0.
\end{eqnarray}
Now, let us expand both gauge fields in plane wave solutions
\begin{align}
 {\bar A}_{\mu}(x)&=\int \frac{d^{2}\mathbf{k}}{\sqrt{\left(2\pi \right)^{3}
 2\bar k_0}}
\,\sum_{\lambda=0 }^2   \varepsilon^{(\lambda)} _{\mu}( \mathbf{k}) (
\,a_{\lambda }(\mathbf{k})\,
e^{-i(\bar k_0 x_0-  {\bf k}\cdot {\bf x} )}  \nonumber \\
& +a_{\lambda }^{\ast }
(\mathbf{k})\,e^{i (\bar k_0 x_0-  {\bf k} \cdot {\bf x} ) }),
\end{align}
\begin{align}
 {\widetilde B}_{\mu}(x)&=\int \frac{d^{2}\mathbf{k}}{\sqrt{\left(2\pi \right)^{3}
\widetilde {k}_0}}
\,    (
\, \eta _{\mu}(\mathbf{k})  b(\mathbf{k})\,
e^{-i ({\widetilde k}_0 x_0-  {\bf k}\cdot {\bf x} ) } \nonumber \\
&  +\eta^*
 _{\mu}(\mathbf{k}) b^{\ast }(\mathbf{k})\,e^{i (\widetilde k_0 x_0-
  {\bf k}\cdot {\bf x} )}),\label{tilde}
\end{align}
where $\bar k_0= |\mathbf{k|}$ and
$\widetilde k_0= \frac{1}{g}\sqrt{1+g^2 |\mathbf{k|}^2}$ correspond to the massless and massive positive frequencies,
respectively, see (\ref{MODES}).
As in the usual case we are free to choose the massless polarization
vectors $\varepsilon^{(\lambda)}$ to satisfy any normalization condition. Hence, we take
\begin{eqnarray}
\varepsilon ^{(\lambda)}(\mathbf{k}) \cdot
\varepsilon^{(\lambda')}(\mathbf{k})= \eta^{\lambda \lambda'}.
\end{eqnarray}
However, according to Eq. (\ref{SEPEQ2}) the massive polarization vector is
uniquely determined and has necessarily to satisfy the normalization condition
\begin{eqnarray}
 \label{EIJ1}
\eta(\mathbf{k}) \cdot \eta^*(\mathbf{k}) &=&-1.
\end{eqnarray}
One finds for both polarization vectors the relations
\begin{eqnarray} \label{PROPPOL}
  \epsilon ^{\mu \beta \gamma}\varepsilon^{(\lambda')}_{\mu}(\mathbf{k}) \varepsilon^{(\lambda)}_{\gamma}(\mathbf{k}) \epsilon ^{\lambda \lambda^{\prime} \lambda^{\prime \prime}}\varepsilon^{(\lambda^{\prime \prime})}_{\beta}(\mathbf{k}),\nonumber
\\ \epsilon ^{ij}\; \eta_{i}(\mathbf{k})\; \eta^{*}_{j}(\mathbf{k}) =-i g k_0,\qquad 
\eta(\mathbf{k}) \cdot \eta(\mathbf{k}) =0,   \\  \eta(\mathbf{k}) \cdot \mathbf{k}=0, 
 \qquad \eta_{i}(\mathbf{k})=\eta_{i}(-\mathbf{k}).\nonumber
\end{eqnarray}
Let us write the canonical Hamiltonian density in the form
\begin{eqnarray}\label{FIRSTORDERHAM}
 \mathcal H_C&=& -\frac{1}{2}  (\dot A_{\mu}\dot A^{\mu}  -
 A_{\mu}\nabla^2  A^{\mu}  )  +\delta  {\mathcal H}_C,
\end{eqnarray}
where the linear part in $g$ is given by
\begin{eqnarray} \label{CORRECTEDHAM}
\delta \mathcal {H}_C&=&  \frac{g}{2}  \epsilon^{ij}
 \dot A_i\Box A_j-\frac{g}{2} \epsilon^{\mu \beta \gamma}\dot A_{\mu}
\partial_{\beta } \dot A_{\gamma}\nonumber \\
&&+\frac{g}{2}  \epsilon^{\mu \beta \gamma}
  A_{\mu} \nabla^2\partial_{\beta }  A_{\gamma}.
\end{eqnarray}
After a lengthy manipulation of replacing the gauge fields in (\ref{FIRSTORDERHAM}) and using the properties
of the polarization vectors (\ref{PROPPOL}) the mode decomposition of the canonical Hamiltonian is
\begin{align}
 H_C=\int d^{2}\mathbf{k}  
 (\bar k_0[ a^{*(1)} a^{(1)}+
 a^{*(2)} a^{(2)}- a^{*(0)}a^{(0)}
 ] -\widetilde k_0 b^{*} b). \nonumber \\ 
\end{align}
We observe the negative-energy
contribution coming from
the time component of the massless
gauge field in the quantity $a^{*(0)}a^{(0)}$ which may be removed for instance
switching to the physical Coulomb gauge as performed in the usual Maxwell theory.
However, irrespective of which gauge is chosen,
the negative-energy contribution from the massive mode  will persist making the total
energy unbounded from below.
At this level a revision of the degrees of freedom is straightforward, see
\cite{COMMENT4}.
\section{A background review} \label{SECTION2}
In this section, the emphasis will be both to recall the canonical formulation for
higher derivative field theories and to introduce the perturbative
method according to
the Refs. \cite{STRING,PERT-METHOD,PERT} that will be used later on.
\subsection{Hamiltonian formulation for higher derivative field theories}
Let us start with the action
\begin{eqnarray}\label{LAGRAN}
S=\int d^4x \; \mathcal L\left( \phi,\partial_{\mu} \phi,\partial_{\mu}\partial_{\nu}
\phi  , \, \dots \, , \,   \partial_{\mu_1}\partial_{\mu_2}  \dots  \partial_{\mu_N} \phi  \right),\nonumber \\
\end{eqnarray}
which for simplicity we assume to be a function of a scalar field
and to depend on a finite number of $N$ derivatives.
A general derivation allowing for more general types of fields
and infinitely many derivatives can be found in \cite{NOETHER}.

The standard procedure of extremizing the action (\ref{LAGRAN}) leads to the generalized Euler-Lagrange equation
\begin{eqnarray}\label{E-L}
\frac{\partial \mathcal L}{\partial \phi}-\partial_{\mu}
\left(\frac{\partial \mathcal L}{\partial(\partial_{\mu} \phi)}\right)
 + \partial_{\mu} \partial_{\nu}
\left(\frac{\partial \mathcal L}{\partial(\partial_{\mu} \partial_{\nu} \phi)}\right)
- \dots \nonumber \\ +(-1)^N \partial_{\mu_1} \dots\partial_{\mu_N}
\left(\frac{\partial \mathcal L}{\partial(\partial_{\mu_1}\dots \partial_{\mu_N} \phi)}\right)=0,
\end{eqnarray}
and analogously to the generalized energy-momentum tensor 
\begin{widetext}
\begin{eqnarray}
T_{\nu}^{\mu}&=&-\delta_{\nu}^{\mu} \mathcal L + \left(
\frac{\partial \mathcal L}{\partial(\partial_{\mu} \phi)} \right)\, \partial_{\nu}\phi
 - \left[   \partial_{\mu_1}
\left(\frac{\partial \mathcal L}{\partial(\partial_{\mu}
\partial_{\mu_1} \phi)}  \right) \partial_{\nu}\phi
- \left(\frac{\partial \mathcal L}{\partial(\partial_{\mu}
\partial_{\mu_1} \phi)}   \right)\partial_{\mu_1} \partial_{\nu}\phi \right]\nonumber
  \\
&& + \left[  \partial_{\mu_1} \partial_{\mu_2}
\left(\frac{\partial \mathcal L}{\partial(\partial_{\mu_1}
\partial_{\mu_2}\partial_{\mu}  \phi)}\right) \partial_{\nu}\phi  -  \partial_{\mu_1}
\left(\frac{\partial \mathcal L}{\partial(\partial_{\mu_1}\partial_{\mu_2}
\partial_{\mu} \phi)}\right)\partial_{\mu_2} \partial_{\nu}\phi +
\left(\frac{\partial \mathcal L}{\partial(\partial_{\mu_1} \partial_{\mu_2}
\partial_{\mu} \phi)}\right)  \partial_{\mu_1} \partial_{\mu_2} \partial_{\nu} \phi   \right]
\nonumber \\ & \vdots & \nonumber \\
&& \dots +(-1)^{N-1} \left[  \partial_{\mu_1} \dots\partial_{\mu_{N-1}}
\left(\frac{\partial \mathcal L}{\partial(\partial_{\mu_1}\dots
\partial_{\mu_{N-1}}\partial_{\mu} \phi)}\right)\partial_{\nu}\phi -  \partial_{\mu_1} \dots \partial_{\mu_{N-2}}
\left(\frac{\partial \mathcal L}{\partial(\partial_{\mu_1}\dots
\partial_{\mu_{N-1}}\partial_{\mu}  \phi)}\right)  \right. \nonumber \\ && \left.
 \times    \partial_{\mu_{N-1}} \partial_{\nu}\phi
 +\; \dots  \;+ (-1)^{N-1}
\left(\frac{\partial \mathcal L}{\partial(\partial_{\mu_1}\dots
\partial_{\mu_{N-1}}\partial_{\mu} \phi)}\right)\partial_{\mu_1} \dots\partial_{\mu_{N-1}}\partial_{\nu}\phi   \right].\nonumber
\end{eqnarray}
Using the Euler-Lagrange equation together with 
\begin{eqnarray}
\partial_{\nu}\mathcal L=\left(  \frac{\partial \mathcal L} {\partial \phi} \right) \partial_{\nu}\phi  +
\left(\frac{\partial \mathcal L}{\partial(\partial_{\alpha_1}  \phi)}\right)\partial_{\alpha_1} \partial_{\nu}\phi
+\left(\frac{\partial \mathcal L}{\partial( \partial_{\alpha_1} \partial_{\alpha_2} \phi)}
\right) \partial_{\alpha_1}  \partial_{\alpha_2}\partial_{\nu}\phi  \dots
+\left(\frac{\partial \mathcal L}{\partial(\partial_{\alpha_1}\dots \partial_{\alpha_{N-1}} \phi)}\right)  \partial_{\alpha_1} \dots \partial_{\alpha_{N-1}} \partial_{\nu}\phi,\nonumber \\
\end{eqnarray}
\end{widetext}
it can be proven that the energy-momentum tensor
satisfies the conservation equation $\partial_{\mu} T_{\; \nu}^{\mu}=0$,
for a detailed derivation see Ref. \cite{NOETHER}.

The corresponding canonical formulation is given by a phase space of dimension $2N$ per space point, been characterized by $N$
configuration field variables which by introducing the notation can be written as
$Q_{0}(t,\mathbf{x})=\phi(t,\mathbf{x}),  Q_{1}(t,\mathbf{x})=\phi^{(1)}
(t,\mathbf{x}),\dots,  Q_{N-1}(t,\mathbf{x})=\phi^{(N-1)}(t,\mathbf{x})$,
where 
\begin{equation}
\phi^{(n)}(t,\mathbf{x})=\frac{\partial^{n}\phi(t,\mathbf{x})}{\partial t^{n}},
\end{equation}
together with their corresponding conjugate momenta $P_{0}(t,\mathbf{x}), P_{1}(t,\mathbf{x}),\dots,P_{N-1}(t,\mathbf{x})$,
and the canonical Hamiltonian density which is identified with the component $T_0^0$, namely
\vspace{-10pt}
\begin{widetext}
\begin{eqnarray}\label{COMP}
{\mathcal H}_C&=&- \mathcal L +
\frac{\partial \mathcal L}{\partial(\partial_{0} \phi)} \partial_{0}\phi
 - \left[   \partial_{\mu_1}
\left(\frac{\partial \mathcal L}{\partial(\partial_{0}
 \partial_{\mu_1} \phi)}  \right) \partial_{0}\phi
- \left(\frac{\partial \mathcal L}{\partial(\partial_{0}
\partial_{\mu_1} \phi)}   \right)\partial_{\mu_1} \partial_{0}\phi \right]
   \\&& + \left[  \partial_{\mu_1} \partial_{\mu_2}
\left(\frac{\partial \mathcal L}{\partial(\partial_{\mu_1}
\partial_{\mu_2}\partial_0 \phi)}\right) \partial_0 \phi  -  \partial_{\mu_1}
\left(\frac{\partial \mathcal L}{\partial(\partial_{\mu_1}\partial_{\mu_2}
\partial_{0} \phi)}\right)\partial_{\mu_2} \partial_{0}\phi +
\left(\frac{\partial \mathcal L}{\partial(\partial_{\mu_1} \partial_{\mu_2}
\partial_{0} \phi)}\right)  \partial_{\mu_1} \partial_{\mu_2} \partial_{0} \phi   \right]
\nonumber    \\ & \vdots & \nonumber    \nonumber \\
&&\dots +(-1)^{N-1} \left[   \partial_{\mu_1} \dots\partial_{\mu_{N-1}}
\left(\frac{\partial \mathcal L}{\partial(\partial_{\mu_1}\dots
\partial_{\mu_{N-1}}\partial_{0} \phi)}\right)\partial_{0}\phi -  \partial_{\mu_1} \dots \partial_{\mu_{N-2}}
\left(\frac{\partial \mathcal L}{\partial(\partial_{\mu_1}\dots
\partial_{\mu_{N-1}}\partial_{0}  \phi)}\right)  
     \partial_{\mu_{N-1}} \partial_{0}\phi+ \dots 
 \right. \nonumber \\ && \left. + (-1)^{N-1}
\left(\frac{\partial \mathcal L}{\partial(\partial_{\mu_1}\dots
\partial_{\mu_{N-1}}\partial_{0} \phi)}\right)\partial_{\mu_1} \dots\partial_{\mu_{N-1}}\partial_{0}\phi   \right].\nonumber
\end{eqnarray}
\end{widetext}
The strategy to determine the conjugate momenta in the above expression
will be to perform successive \emph{spatial} integration by parts in the fields to subsequently regroup the quantities
multiplying the pure 
time-derivative terms $\partial^n_0 \phi$. 
Therefore, after successive integration by parts, the canonical Hamiltonian density (\ref{COMP}) can be given in terms of space phase variables as
\begin{eqnarray}
{\mathcal H}_C&=&P_0(x) \dot Q_0(x)+ P_1(x) \dot Q_1(x)  +\dots \nonumber \\
&&+P_{N-1}(x)\dot Q_{N-1}(x) - \mathcal L,
\end{eqnarray}
from we read the explicit expressions for conjugate momenta
\begin{eqnarray}
 P_{N-1}(t,\mathbf{x})&=&\frac{\partial \mathcal L}{\partial Q_{N}},
\end{eqnarray}
\begin{eqnarray}\label{HIGHERMOMENTA}
 P_{i}(t,\mathbf{x})&=&\frac{\partial \mathcal L}{\partial Q_{i+1}}
 +\sum_{j=1}^{N-i-1} \frac{(-1)^j(j+i+1)!}{j!\,(i+1)!}\;  \partial_{m_1}\dots \partial_{m_j}\nonumber \\
 && \times
\left(\frac{\partial \mathcal L}{\partial \partial_{m_1}
\dots \partial_{m_j}Q_{i+1}}\right)
-  \frac{\partial P_{i+1}}{\partial t} ,
\end{eqnarray}
the index $i$ runs trough $i=0,\dots,N-2$ and $\partial_{m_k}$ stands for spatial derivatives.
We have the compact expressions for the canonical Hamiltonian and the generalized symplectic form
\begin{eqnarray}
 H_C&=&\int d^{3}\mathbf{x\;} \left( \sum_{i=0}^{N-1}\;P_{i}\dot Q_{i}-\;\mathcal{L}
\right),
\end{eqnarray}
\begin{eqnarray}
\Omega (t)&=&\int d^{3}\mathbf{x\;} \left(  \sum ^{N-1}_{i=0}  dP_{i}
\left(\mathbf{x},t\right) \wedge d Q_i\left( \mathbf{x},t\right) \right).
\end{eqnarray}
The basic Poisson brackets which can be read off from the above symplectic form are
\begin{eqnarray}
\left\{ Q_i(t,\mathbf{x}),P_{j}
(t,\mathbf{x}^{\prime})
\right\}&=&\delta
_{ij}\delta ^{3}(\mathbf{x}-
\mathbf{x}^{\prime}), \nonumber \\
 i,j&=&0,\dots ,N-1.
\end{eqnarray}
Let us remark that due to possible 
degeneracy of the higher derivative theory, there may not be a unique
solution expressing
$  Q_{i}\left( \mathbf{x},t\right)$ in terms of the other canonical variables, occurring when
$\left(\frac{\partial^2 \mathcal L}{\partial Q_N\partial Q_N}\right)=0$.
In this case one can follow the Hamiltonian treatment for singular higher derivative systems developed in Ref. \cite{PONS}.
\subsection{The perturbative method}
Let us consider a framework of a system having a Lagrangian density of the form
\begin{equation}
\mathcal{L}=\mathcal{L}_0(\phi \, , \, \partial_\mu{\phi})
+g\, \mathcal L_1 ,
\end{equation}
where $\mathcal L_1$ is an arbitrary function of the fields (not necessarily quadratic) that
contains all the higher order derivative dependence.
In the limit
of the coupling parameter $g \to 0$
we recover the ordinary derivative theory defined by $\mathcal L_0$.

The perturbative method for a canonically formulated higher derivative theory
can be described in the following steps:

(i) In order to obtain the appropriate Hamiltonian to order $g^{p}$ one
starts by iteratively solving the equation of motion to the order
$g^{p-1}$.

(ii) Using the above iterated equation we express all the time-derivatives of the fields $\phi^{(q)}(t,\mathbf{x})$
with $q > 2$ in terms of the first-order variables
$\phi(t,\mathbf{x}), \dot \phi(t,\mathbf{x})$,
called from now velocity phase space variables, and their possible spatial derivatives.
This will introduce
further contributions in powers of the perturbation parameter $g$
which need to be maintained only up to the required order.

(iii) We substitute the above expressions
$\phi ^{(q)}(t,\mathbf{x}),\,\, q > 2$ 
into the conjugate momenta $P_j$, the canonical Hamiltonian and the
generalized symplectic form $\Omega(t)$. This allows to recast up to the order $p$ in terms
of velocity phase space variables the first-order expressions
\begin{equation}
P_j^{(p)}=P^{(p)}_j(\phi,\dot \phi)+\mathcal O(g^{p+1}),
\end{equation}
\begin{equation}
{H}^{(p)}=\int d^3 {\bf x} \;  \mathcal{H}^ {(p)} (\phi, \, \dot{\phi} )+\mathcal O(g^{p+1}),
\end{equation}
\begin{eqnarray}\label{APPROXSYMFORM}
 \Omega^{(p)}(t)&=&\iint 
d^3 {\bf x} \ d^3{\bf y}  \hat \Omega^{(p)}\left(t, \mathbf{x} , \mathbf{y} \right) \nonumber \\
&& \times \left( d
\dot{\phi}(t, \mathbf{x})\, \wedge\, d \phi(t, \mathbf{y})\right)+\mathcal O(g^{p+1}),
\end{eqnarray}
From above we can deduce the Poisson bracket
\begin{equation}
 \{ \phi(t, \mathbf{x}) \, , \, \dot{\phi}%
(t, \mathbf{y})\}=( \hat \Omega^{(p)}\left(t, \mathbf{x},\mathbf{y} \right))^{-1},
\end{equation}
the hat in $\hat \Omega^{(p)}\left(t, \mathbf{x} , \mathbf{y} \right)$ denotes possible dependence on spatial derivatives.

(iv) In order to diagonalize (\ref{APPROXSYMFORM}), we search for an invertible change of
variables from velocity phase variables $( \phi (t,\mathbf{x}),\dot{\phi}(t,\mathbf{x}))$ to new
canonical ones  $( \widetilde Q(\phi
,\dot{\phi}),\;\widetilde P(\phi ,\dot{\phi}))$
in such a way that the
new Poisson bracket $\{\widetilde Q(t,\mathbf{x})\, , \, \widetilde P(t,%
\mathbf{y})\}$ are canonical to the order considered. That is to
say
\begin{eqnarray}
\{\widetilde Q(t,\mathbf{x})\, , \, \widetilde P(t,%
\mathbf{y })\}&=& \int d^3z \ \ d^3z^{\prime}\left(\frac{\delta
\widetilde Q(t,\mathbf{x})}{\delta
\phi(t,\mathbf{z})}\frac{\delta \widetilde P(t,\mathbf{y })}{\delta \dot{\phi}(t,\mathbf{%
z^{\prime}})} \right. \nonumber \\ && -\left.\frac{\delta \widetilde Q(t,\mathbf{x})}{ \delta \dot{\phi}(t,\mathbf{%
z^{\prime})}} \frac{\delta \widetilde P(t,\mathbf{y })}{\delta \phi(t,\mathbf{z})}%
\right) \nonumber \\
&& \times \{ \phi(t,\mathbf{z}) \, , \,
\dot{\phi}(t,\mathbf{z^{\prime}})\}    \notag \\
&&=           \delta^3 ({\mathbf x}- {\mathbf y}) +\mathcal O(g^{p+1})       .
\end{eqnarray}

(v) Having the kinematical theory completed we proceed with the dynamics of the theory.
Thus, replacing the new variables we rewrite the Hamiltonian as
\begin{equation}
{ H}^{(p)}=\int d^3 {\bf x} \;  \mathcal{H}^ {(p)} (\phi(  \widetilde Q,\widetilde P), \dot \phi( \widetilde Q,\widetilde P ) ),
\end{equation}
and using the Hamilton equation
\begin{eqnarray}
\dot {\widetilde Q}=\left\{ \widetilde Q,
H^{(p)}\right\},
\end{eqnarray}
we write conjugate momenta $\widetilde P$ in terms of  $\widetilde Q$ and $\dot {\widetilde Q}$.
Finally, by using the Legendre transformation
we arrive to the effective Lagrangian density
\begin{eqnarray}
\tilde{\mathcal{L}}= \widetilde P(Q ,\dot {\widetilde Q})\, \dot{\widetilde Q}-{%
\mathcal{H}}(\widetilde Q,\dot {\widetilde Q}).
\end{eqnarray}
Therefore, in terms of the new variables $\widetilde Q,\widetilde  P$ both the Hamiltonian 
density ${\mathcal{H}}(\widetilde Q ,\widetilde P),$
together with the Poisson bracket $\left\{\widetilde  Q (t,\mathbf{x}), \widetilde  
P(t,\mathbf{y })\right\}=\delta^3(\mathbf{x}-\mathbf{y })$
define the physical approximation of the system to the order
considered.
The first-order Hamiltonian will be bounded from below provided the initial
one obtained from $\mathcal{L}_0$ is.
One can check that the Euler-Lagrange equations reproduce those of the original system to the order considered;
a proof of self-consistency to all orders is
provided in a mechanical
setting in Ref. \cite{METHODPERT}.
\section{The effective theory }
In this section we describe the main ingredients and results of the application of the
perturbative method to the extended MCS theory \cite{COMMENT}.
The method has been previously applied for fermions and
scalars in Refs. \cite{METHODPERT,REYES}.
\vspace{-10pt}
\subsection{The reduced phase space}
We start by iteratively solving the
equation of motion (\ref{EQ}) to the lowest order approximation ($p=1$). A first iteration gives
\begin{equation}
 \Box  A^{\sigma}=g^2( 
\partial^{\sigma}\Box (\partial \cdot A)-\Box^2A^{\sigma}     )\approx \mathcal O(g^2).
\end{equation}
Therefore, to the order considered, our iterated equation of 
motion is the basic equation
\begin{eqnarray}\label{ITERATION}
 \ddot A_{\alpha} = \nabla^2  A_{\alpha}+\mathcal O(g^2).
\end{eqnarray}
Replacing into the set of canonical variables (\ref{MOM1}) we arrive to
\begin{eqnarray}\label{ITERATION1}
 && P^i=F_{0i}+\frac{g}{2}
\epsilon^{ij} \nabla^2 A_j
 -\frac{g}{2} \epsilon^{ij} \partial_j \dot A_{0}+\mathcal O(g^2),
\\&&
P^0= -  (\partial \cdot A)
-\frac{g}{2} \epsilon^{ij} \partial_i
\dot A_j,\\
&& \Pi^0=\frac{g}{2} \epsilon^{ij}
\partial_iA_j,
\\ &&   \Pi^i= -\frac{g}{2} \epsilon^{ij} F_{0j}.
\end{eqnarray}
The perturbed Hamiltonian density (\ref{FIRSTORDERHAM}) gets expressed by
\begin{eqnarray}\label{FIRSTORDERHAM2}
  {\mathcal { H}^{(1)}}&=& -\frac{1}{2}(\dot A_0^2 -
 A_0\nabla^2 A_0) +\frac{1}{2}(\dot A_i^2-
 A_i\nabla^2A_i)\nonumber \\&& - g\epsilon^{ij}
\partial_i \dot A_j\dot A_0+g \epsilon^{ij} \nabla^2A_j\dot A_i
+g\epsilon^{ij} \partial_i A_j\nabla^2 A_0\nonumber
\\ &&+\mathcal O(g^2),
\end{eqnarray}
and the perturbed symplectic form
\begin{align}\label{FORMRED}
  \Omega^{(1)}(t) &= \int d^{2}\mathbf{x\;}
  \left( d \dot A_i \wedge d A_i-d\dot A_0
\wedge d  A_0  \right. \nonumber \\
&\left. - g\epsilon^{ij} [d( \partial_i
\dot A_j)\wedge d  A_0
+ d( \partial_j \dot A_0)\wedge d  A_i  ]
\right. \nonumber \\
& \left. +\frac{g}{2}
\epsilon^{ij} [d \dot A_i \wedge d \dot A_j
- d( \nabla^2 A_i) \wedge d A_j ]   \right)+\mathcal O(g^2).\nonumber \\
\end{align}
Let us consider the equivalent two-point split expression
\begin{align} \label{SYMP-FORM}
 \Omega^{(1)} (t)=\frac{1}{2}\iint d^{2}\mathbf{x\;}
d^{2}\mathbf{x}^{\prime
}    \hat \Omega^{(1)}_{ab}(t;\mathbf{x,x}^{\prime })
\;dz^{a}(t,\mathbf{x})\wedge dz^{b}(t,%
\mathbf{x}^{\prime }),\nonumber \\
\end{align}
where
\begin{widetext}
\begin{equation} \label{SIMPMATRIX}
 \hat \Omega^{(1)}_{ab}(t;\mathbf{x,x}^{\prime })=\left[
\begin{array}{cccccc}
0 & 0 & 0 & 1& -g \partial_2 & g \partial_1  \\
0 & 0 & -g\nabla^2 & g\partial_2  & -1 & 0\\
0 & g\nabla^2 & 0 & -g\partial_1 & 0 & -1 \\
-1 & g\partial_2 & -g\partial_1 & 0 & 0 & 0 \\
-g\partial_2  & 1 & 0 & 0&0&  g\\
g\partial_1 & 0& 1 & 0 & -g & 0
\end{array}%
\right] \delta ^{3}(\mathbf{x}-\mathbf{x}^{\prime }),
\end{equation}
and the notation is such that $z^a=(A^{0},A^{1},A^{2},\dot A^{0}, \dot A^{1},
\dot A^{2}  )$ with $a=1,2,3,4,5,6$.
Notice that a minus sign has appeared in the
linear derivatives due to an integration by parts. Also,
here and in the following, to avoid ambiguities we will consider the action of the derivatives with respect to
the unprimed coordinate $\mathbf{x}$.

The inverse symplectic matrix is easily
computed to give
\begin{equation}\label{INV}
(\hat {\Omega}^{(1)}_{ab} (t;\mathbf{x,x}^{\prime }))^{-1} =\left[
\begin{array}{cccccc}
0 & 0 & 0 & -1& g \partial_2 & -g \partial_1 \\
0& 0 & g  & -g \partial_2 & 1 & 0\\
0 & -g  & 0 & g \partial_1 & 0 & 1  \\
1  &  -g \partial_2 & g \partial_1  & 0 & 0 & 0 \\
g  \partial_2 & -1 & 0 & 0 & 0 & -g \nabla^2 \\
-g \partial_1 & 0& -1 & 0& g \nabla^2 & 0
\end{array}%
\right] \delta ^{3}(\mathbf{x}-\mathbf{x}^{\prime }).
\end{equation}
\end{widetext}
It can be checked that the usual properties for the matrices (\ref{SIMPMATRIX}) and (\ref{INV}),
such as the antisymmetry property
\begin{eqnarray}\label{MATRIXW}
\hat \Omega^{(1)}_{ab}
(t;\mathbf{x,x}^{\prime })=-
\hat \Omega^{(1)}_{ba}(t;\mathbf{x}^{\prime }
\mathbf{,x}),
\end{eqnarray}
together with the inverse relation
\begin{align}
\int d^{2}\mathbf{x}^{\prime }\;
\hat \Omega^{(1)}_{ab}(t;\mathbf{x,x}^{\prime })
(\hat \Omega^{(1)}_{bc}(t;
\mathbf{x}^{\prime },\mathbf{x}^{\prime
\prime }))^{-1}=\delta _{ac}\;\delta
^{3}(\mathbf{x-x}^{\prime \prime }),\nonumber \\
\end{align}
are satisfied up to linear order in $g$.

Now, considering that
\begin{equation}\label{POISSON}
\left\{ z^{a}(t,\mathbf{x}),\;z^{b}
(t,\mathbf{x}^{\prime })\right\}
= (\hat \Omega^{(1)}_{ab}(t;\mathbf{x,x}^{\prime }))^{-1},
\end{equation}
 from (\ref{INV}) the nonzero brackets are
\begin{eqnarray}\label{PERTPOISSON}
\{ A_{0}({t,\mathbf{x} }),\dot A_{0}({ t,\mathbf{x}}^{\prime}) \}&=&-
\delta ^{3}(\mathbf{x}-\mathbf{x}^{\prime }),\nonumber
\\
\{ A_{i}({ t,\mathbf{x}}),\dot A_{j}({ t,\mathbf{x}}^{\prime}) \}&=&
\delta _{ij}\;\delta ^{3}(\mathbf{x}-\mathbf{x}^{\prime }),\nonumber
\\
 \{A_i({t,\mathbf{x}} ), A_j({t,\mathbf{x}}^{\prime}) \}&=&g\epsilon^{ij}\;
\delta ^{3}(\mathbf{x}-\mathbf{x}^{\prime }),\nonumber
\\
\{  A_i({ t,\mathbf{x}}),\dot A_0({t,\mathbf{x} }^{\prime}) \}&=&-g
\epsilon^{ij} \frac{\partial }
{\partial {  x}^j}\; \delta ^{3}(\mathbf{x}-\mathbf{x}^{\prime }),\nonumber
\\
\{A_0({t,\mathbf{x}}),\dot A_i ({t,\mathbf{x}}^{\prime})\}&=&g\epsilon^{ij}
\frac{\partial }
{\partial { x}^j}\; \delta ^{3}(\mathbf{x}-\mathbf{x}^{\prime }),
\nonumber
\\
 \{  \dot  A_i({t,\mathbf{x}}),\dot A_j({t,\mathbf{x}}^{\prime}) \}
 &=&-g\epsilon^{ij} \nabla^2\; \delta ^{3}(\mathbf{x}-\mathbf{x}^{\prime }).
\end{eqnarray}
\subsection{New canonical variables}
Until now we have the effective theory given in terms of velocity phase space variables with their corresponding Poisson brackets
exhibiting the non standard form (\ref{PERTPOISSON}).
According to the method we need to search for a new set of canonical variables
for which the Poisson brackets (\ref{PERTPOISSON}) are diagonal. The effective theory will be constructed in terms of these variables.
Therefore, let us consider the invertible change of variables
\begin{eqnarray}
 A &\rightarrow& \widetilde A ( A,\dot A), 
\end{eqnarray}
\begin{eqnarray}
 \dot A &\rightarrow& \widetilde \pi ( A,\dot A),
\end{eqnarray}
such that (\ref{FORMRED}) can be put in the form
\begin{eqnarray}\label{SYMPLECTICAPPROX}
\Omega^{(1)} (t)=\int d^{2}\mathbf{x\;}
d\widetilde \pi^{\mu}(t,\mathbf{x})
\wedge d { {\widetilde  A}}_{\mu}(t,\mathbf{x})+\mathcal O(g^2),
\end{eqnarray}
or alternatively
\begin{eqnarray}\label{POISSONAPPROX}
&&\left\{  \widetilde A_{\mu}\left(A({ t,\bf{x}}),\dot A({ t,\bf{x}})\right),
 \widetilde \pi_{\nu}\left(A({ t,\bf{y}}),\dot
  A({ t,{\bf y}})\right)\right\}\nonumber \\
  &&=\eta_{\mu \nu}\delta ^{3}(\mathbf{x}-
\mathbf{y})+\mathcal O(g^2),
\end{eqnarray}
which may be verified with the use of (\ref{PERTPOISSON}).
It is important to mention that the above realization can be achieved by more than one set of canonical variables.
Hence, in order to fix a unique set of canonical variables we will impose additional criteria on the gauge field redefinition:
These requirements are: (i) we choose the new gauge field $\tilde A^{\mu}$ to transform covariantly under Lorentz transformations and
(ii) to satisfy the same gauge fixing condition as the old one.

Here the simplest way to proceed is to select a new set of canonical variables
and check if Eqs. (\ref{SYMPLECTICAPPROX}), (\ref{POISSONAPPROX}) together
with the two above conditions are verified. Therefore, let us consider the twisted gauge redefinition
\begin{eqnarray}\label{NEWVAR}
 \widetilde A ^{\mu}= A^{\mu}+ \frac{g}{2}\epsilon^{\mu \alpha \beta}
 \partial_ {\alpha }A_{\beta},
\end{eqnarray}
which in components becomes
\begin{eqnarray}
\widetilde A _i&=& A_i+ \frac{g}{2}\epsilon^{ij} F_{0j},
\\
 \widetilde A _0&=& A_0+\frac{g}{2}\epsilon^{ij} \partial_iA_j.
\end{eqnarray}
We define the corresponding canonical momenta by
\begin{eqnarray}
 { \widetilde \pi_{0}}= -(\partial\cdot A),
\end{eqnarray}
\begin{eqnarray}
{\widetilde \pi^{i}}= F_{0i}+ \frac{g}{2}
 \epsilon^{ij} \nabla^2 A_j
  -  \frac{g}{2} \epsilon^{ij} \partial_j \dot A_0-
  \frac{g}{2}\epsilon^{lm} \partial_i\partial_l  A_m.
\end{eqnarray}
One can check that the new gauge field respects the gauge fixing condition
\begin{eqnarray}
\chi_{G.F}= (\partial \cdot A)(\partial \cdot \widetilde A),
\end{eqnarray}
and transforms covariantly under Lorentz
transformations, which we further prove.

It is not a difficult task to check that the new variables diagonalize the Poisson brackets,
using (\ref{PERTPOISSON}) let us compute the Poisson brackets
\begin{align}
\{ \widetilde A _0({t,\mathbf{x} }) , \widetilde \pi_{0} ({t,\mathbf{x}^{\prime} })
 \}&=\left\{  A_0+\frac{g}{2}\epsilon^{ij} \partial_iA_j  ,-(\partial\cdot A) \right\}\nonumber
\\ &=-\left\{  A_0+\frac{g}{2}\epsilon^{ij} \partial_iA_j  ,\dot A_0-\partial _k A_k \right\} \nonumber \\
&=\delta ^{3}(\mathbf{x}-\mathbf{x}^{\prime })+\mathcal O(g^2),
\end{align}

\begin{align}
\{  \widetilde A _i({t,\mathbf{x} }) , \widetilde \pi ^j({t,\mathbf{x} }^{\prime}) \}&=\Big\{A_i+ \frac{g}{2}
 \epsilon^{im} F_{0m}
, F_{0j}+ \frac{g}{2}\epsilon^{jk}\nabla^2 A_k\nonumber \\ &
- \frac{g}{2}\epsilon^{jm} \partial_m \dot A_0-
  \frac{g}{2} \epsilon^{lm} \partial_j\partial_l  A_m  \Big\}\nonumber \\
  &=\eta_i ^{j} \delta ^{3}(\mathbf{x}-\mathbf{x}^{\prime })+\mathcal O(g^2),
\end{align}
and 
\begin{align}
\{  \widetilde A_i({t,\mathbf{x} }), \widetilde A_j({t,\mathbf{x}^{\prime} }) \}&=\Big\{A_i+ \frac{g}{2} \epsilon^{im}
F_{0m} , A_j+  \frac{g}{2} \epsilon^{jm} F_{0m} \Big\}\nonumber \\ &=g\epsilon^ {ij}
  +\frac{g}{2}\epsilon^ {jm} \delta_{im} -\frac{g}{2}
  \epsilon^ {im} \delta_{mj} +\mathcal O(g^2)\nonumber \\
  &=0,
\end{align}
from where we finally arrive to
\begin{eqnarray}\label{DIAGFORM}
\{  \widetilde A_{\mu}({t,\bf x}),
{ \widetilde \pi_{\nu}({t,\bf x^{\prime}})
}\}&=&\eta_{\mu \nu}\delta ^{3}(\mathbf{x}-
\mathbf{x}^{\prime })+\mathcal O(g^2).
\end{eqnarray}
The same result follows by considering
the inverse change of variables
\begin{eqnarray}\label{T1}
  A_0&=& \widetilde A_0- \frac{g}{2}\epsilon^{ij}
  \partial_i \widetilde A_j,\nonumber
\\ \label{T2}
  A_i&=& \widetilde A _i-\frac{g}{2}\epsilon^{ij}
  \widetilde \pi^ j,\nonumber
\\
  \dot A_0&=&\partial_k  \widetilde A_k - {\widetilde \pi_0}-
  \frac{g}{2} \epsilon^{ij} \partial_i \widetilde \pi^j,\nonumber
\\ \dot A_i&=&\widetilde \pi^i+ \partial_i \widetilde A_0 +
  \frac{g}{2}\epsilon^{ij} \partial_j \partial_k \widetilde A_k
 - \frac{g}{2}\epsilon^{ij} \partial_j  \widetilde \pi_0-\frac{g}{2}
 \epsilon^{ij}\nabla^2 \widetilde A_j,\nonumber \\
\end{eqnarray}
and replacing in the perturbed symplectic form
\begin{eqnarray}
  \Omega^{(1)}(t) &=& \int d^{2}\mathbf{x\;}
  ( d \dot A_i \wedge d A_i-d\dot A_0
\wedge d  A_0 \nonumber \\
&& -g\epsilon^{ij} [d( \partial_i
\dot A_j)\wedge d  A_0
+ d( \partial_j \dot A_0)\wedge d  A_i ]
\nonumber \\
&& +\frac{g}{2}
\epsilon^{ij} [d \dot A_i \wedge d \dot A_j
- d( \nabla^2 A_i) \wedge d A_j ]  ),
\end{eqnarray}
which after some algebra produces
\begin{eqnarray}
 \Omega^{(1)}(t)&=&\int d^{2}\mathbf{x\;\;} d \widetilde \pi_{\mu}
  (t,{\bf x})\wedge d  \widetilde A^{\mu}(t,{\bf x}),
\end{eqnarray}
in agreement with the result (\ref{DIAGFORM}).
\subsection{The low-energy effective Lagrangian}
Recall from the Eq. (\ref{FIRSTORDERHAM2}) the canonical Hamiltonian density written in the old variables
\begin{eqnarray}
  {\mathcal { H}^{(1)}}&=& -\frac{1}{2}\dot A_0^2 +\frac{1}{2}
 A_0\nabla^2 A_0  +\frac{1}{2}\dot A_i^2-\frac{1}{2}
 A_i\nabla^2A_i \nonumber \\&& - g\epsilon^{ij}
\partial_i \dot A_j\dot A_0+g \epsilon^{ij} \nabla^2A_j\dot A_i
+g\epsilon^{ij} \partial_i A_j\nabla^2 A_0,\nonumber \\
\end{eqnarray}
which with the use of the inverse transformations has to be expressed in terms of the new canonical variables.
Using the inverse transformations (\ref{T1}) we have the expressions
\begin{align}
-\frac{1}{2} \dot A_0^2& =-\frac{1}{2}(\partial_k \widetilde A_k)^2-\frac{1}{2}(\widetilde \pi_0)^2+\widetilde \pi_0(\partial_k \widetilde A_k)
\nonumber \\ & +\frac{g}{2}\epsilon^{ij}\partial_i \widetilde \pi^j(\partial_k \widetilde A_k-\widetilde \pi_0),
\nonumber \\ -\frac{1}{2} (\partial_i A_0)^2 &= - \frac{1}{2} (\partial_i \widetilde  A_0)^2
+\frac{g}{2}\epsilon^{lm}(\partial_i  \partial_l  \widetilde A_m )(\partial_i \widetilde A_0), \nonumber \\
\frac{1}{2} \dot A_i^2 &=\frac{1}{2} (\widetilde \pi^i)^2+\frac{1}{2}(\partial _i\widetilde A_0)^2+\widetilde \pi^i(\partial_i \widetilde A_0)
\nonumber \\ &+  \widetilde \pi^i (\frac{g}{2}\epsilon^{ij}\partial_j \partial_k \widetilde A_k-\frac{g}{2}\epsilon^{ij}\partial_j \widetilde \pi_0-\frac{g}{2}\epsilon^{ij}\nabla^2 \widetilde A_j)\nonumber \\
&-\frac{g}{2}\epsilon^{ij}\partial_i   \widetilde A_0 \nabla^2 \widetilde A_j,
\nonumber \\
\frac{1}{2} (\partial_i A_j)^2 &=  \frac{1}{2} (\partial_i \widetilde  A_j)^2
-\frac{g}{2}\epsilon^{jk}\partial_i  \widetilde \pi^k \partial_i \widetilde A_j,
\end{align}
and the ones linear in $g$
\begin{eqnarray}
 - g\epsilon^{ij}
\partial_i \dot A_j\dot A_0&=& -g \epsilon^{ij}\partial_i  \widetilde \pi^j (\partial_k \widetilde A_k)
+g \epsilon^{ij}(\partial_i  \widetilde \pi^j) \widetilde\pi_0,
\nonumber \\
g \epsilon^{ij} \nabla^2A_j\dot A_i&=&g \epsilon^{ij}\widetilde \pi^i  \,\nabla^2  \widetilde A_j
+g \epsilon^{ij}\nabla^2  \widetilde A_j (\partial_i \widetilde A_0),
\nonumber \\
g\epsilon^{ij} \partial_i A_j\nabla^2 A_0&=&g \epsilon^{ij}(\partial_i  \widetilde A_j) \nabla^2\widetilde A_0.
\end{eqnarray}
Replacing in the Hamiltonian, we arrive to
\begin{eqnarray}
 \mathcal{  H}^{(1)}&=&-\frac{1}{2}
 \widetilde \pi_0^2+ \frac{1}{2} (\widetilde \pi^i)^2
+\widetilde \pi_0 (\partial_k  \widetilde A_k ) -
  \widetilde A_0 (\partial_i \widetilde \pi^i )\nonumber \\ &&+\frac{1}{4}
  F_{ij}^2(\widetilde A),
\end{eqnarray}
where $F_{ij}(\widetilde A)=\partial_i\widetilde A_j
-\partial_j\widetilde A_i$.

Considering the Hamilton equation $\dot {\widetilde A_{\mu}}=\left\{
\widetilde A_{\mu},H^{(1)}\right\}$ we obtain the canonical momenta in terms of configuration variables
and their first time derivatives  
\begin{eqnarray}\label{DEFMOMNEW}
   \widetilde \pi_0 =-(\partial \cdot \widetilde A),\qquad
  \widetilde \pi^i=\dot {\widetilde A_{i}}-
   \partial_i\widetilde A_0.
\end{eqnarray}

The effective Lagrangian is obtained via the Legendre transformation
\begin{eqnarray}
 \widetilde{\mathcal   L}
=\widetilde \pi ^ {\mu}   \dot {\widetilde A}_
{\mu}-{\mathcal H}^{(1)}( \widetilde A,\widetilde \pi).
\end{eqnarray}
Finally, using the relations (\ref{DEFMOMNEW}) we obtain
\begin{eqnarray} \label{EFFECTIVEL}
 \widetilde{\mathcal   L}
=-\frac{1}{4} F_{\mu \nu }(\widetilde A)F^{\mu \nu }(\widetilde A)
-\frac{1}{2}(\partial \cdot \widetilde A)^2.
\end{eqnarray}
Therefore, considering that
$\widetilde A_{\mu}= A_{\mu}+
\frac{g}{2}\epsilon_{\mu
\alpha \beta} \,\partial^{\alpha }A^{\beta}$
and $F_{\mu \nu}(\widetilde A)\partial_{\mu}\widetilde A _{\nu}- \partial_{\nu}
\widetilde A _{\mu} $ the Maxwell theory is correctly reproduced by setting $g\to 0$.

The standard variation produces
the equation of motion
\begin{equation}
 \Box \widetilde A_{\alpha}=0,
\end{equation}
and by considering plane waves with respect to the new gauge field gives
the solution $k^ 2=0$. We see that only the massless
 solution has been retained
in the effective field theory while the massive
mode has been removed.
\subsubsection{Gauge invariance}
Recall that classically appropriate conditions can be
imposed on the boundary data in order to set
$\partial \cdot \widetilde A=0$ everywhere so we can lift
this condition from the effective theory. 
Therefore, let us consider the Lagrangian
\begin{eqnarray}
\widetilde {\mathcal   L}
=-\frac{1}{4}F_{\mu \nu }(\widetilde A) F^{\mu \nu }(\widetilde A),
\end{eqnarray}
which is manifestly invariant under the gauge transformation ${\widetilde A}^{\prime}_{\mu}= {\widetilde A}_{\mu}-\partial_{\mu}\Lambda$.
\subsubsection{Lorentz symmetry}
Lorentz invariance follows naturally in the effective theory by proving covariance of the
gauge field $\widetilde A^{\mu}$. Therefore, consider the Lorentz transformation
\begin{eqnarray}
\widetilde A^{\mu}\to  \widetilde A^{\mu^\prime}&=& A^{ \mu^\prime}+ \frac{g}{2}
 \epsilon^{\mu^\prime \nu^\prime \rho^\prime} \partial_ { \nu ^\prime}A_{ \rho^\prime},
\end{eqnarray}
and replace the transformation for the original fields 
\begin{eqnarray}
A^{\mu^\prime}=\Lambda^{\mu^\prime} _{\;\alpha } \,A^{\alpha},
\end{eqnarray}
together with the relation $\epsilon^{\mu^\prime \nu^\prime \rho^\prime } \: \Lambda_{\; \nu^\prime}^{ \sigma} \Lambda_{\rho^\prime}^
 {\; \lambda}=\epsilon^{\alpha \sigma \lambda}  \Lambda^{\mu^\prime}_{\; \alpha}$. This
gives the desired relation
\begin{eqnarray}
 \widetilde A^{ \mu^\prime}&=&\Lambda
 _{\; \alpha}^{\mu^\prime}  \left(A^{\alpha}
 + \frac{g}{2}
 \epsilon^{\alpha \sigma \lambda}
\, \partial_{\sigma }A_{\lambda}\right)
  =\Lambda
 _{\;\alpha}^{\mu^\prime} \; \widetilde A^{ \alpha}.
\end{eqnarray}
\subsubsection{$\mathcal C \mathcal P\mathcal T $ symmetries}
Parity transformation $\mathcal P$ is defined in $2+1$ dimensions by
\begin{eqnarray}\label{PARITY}
 x^0 \rightarrow  x^0, \qquad
 x^1 \rightarrow  -x^1, \qquad x^2\rightarrow x^2,
\end{eqnarray}
which corresponds to a reflection in just one of the spatial axes \cite{DESER2}.
Notice that this transformation yields the improper transformation defined to have $\det \Lambda=-1$, instead
of the space inversion
${\bf x} \to -{\bf x}$ one is familiar in three spatial dimensions.

The original gauge field transforms as
\begin{eqnarray}\label{PARITYFIELD}
 \mathcal P A ^0(x^0,{\bf x})  \mathcal P^{-1} &=&  A^0(x^0,{\bf x}^{\prime}),\nonumber
\\
  \mathcal P A^1(x^0,{\bf x})\mathcal P^{-1} &=&  -A^1(x^0,{\bf x}^{\prime}),\nonumber
\\
   \mathcal P A^2(x^0,{\bf x})\mathcal P^{-1} &=&  A^2(x^0,{\bf x}^{\prime}),
\end{eqnarray}
where ${\bf x}=(x^1,x^2)$ and ${\bf x}^{\prime}=(-x^1,x^2)$.

From (\ref{NEWVAR}) we have in components
\begin{eqnarray}\label{COMPNEW}
 \widetilde A^0(x^0,{\bf x})= A^0(x^0,{\bf x})+\frac{g}{2}\left(\partial_1A_2(x^0,{\bf x})-\partial_2A_1(x^0,{\bf x})\right),\nonumber \\
 \widetilde A^1(x^0,{\bf x})=A^1(x^0,{\bf x})+\frac{g}{2}\left(\partial_2A_0(x^0,{\bf x})-\partial_0A_2(x^0,{\bf x})\right),\nonumber \\
 \widetilde A^2(x^0,{\bf x})=A^2(x^0,{\bf x})+\frac{g}{2}\left(\partial_0A_1(x^0,{\bf x})-\partial_1A_0(x^0,{\bf x})\right).\nonumber \\
\end{eqnarray}
In order to find expressions analogous to those of Eqs. (\ref{PARITYFIELD}) it is convenient to set 
$A_{\mu}(x)=\epsilon_{\mu}(k)e^{-ik \cdot x}$, where Eq. (\ref{COMPNEW}) becomes
\begin{eqnarray}\label{TRANSF}
 \widetilde A^{\mu}(x^0,{\bf x}) &=&  T^{\mu}_{ \;\alpha}(k^0,{\bf k} ) A^{\alpha}(x^0,{\bf x}),
\end{eqnarray}
such that
\begin{equation}
 T(k^0,{\bf k} ) =\left[
\begin{array}{cccccc}
1 & -igk_2/2 &  igk_1/2  \\
-igk_2/2 & 1 &-igk_0/2 \\
 igk_1/2& igk_0/2 & 1
\end{array}%
\right].
\end{equation}
Under a parity transformation we find  
\begin{eqnarray}\label{PARITYTRANS}
 \mathcal P\, \widetilde A^{\mu}(x^0,{\bf x}) \, \mathcal P^{-1}&=&
 R^{\mu}_{ \;\alpha}(k^0,{\bf k} ) A^{\alpha}(x^0,{\bf x}^{\prime}),
\end{eqnarray}
with
\begin{equation}
 R(k^0,{\bf k} ) =\left[
\begin{array}{cccccc}
1 & igk_2/2 &  -igk_1/2  \\
-igk_2/2 & -1 &-igk_0/2 \\
 -igk_1/2& -igk_0/2 & 1
\end{array}%
\right].
\end{equation}
We need the original gauge field evaluated at ${\bf x}^{\prime}$, so from (\ref{TRANSF}) we get
\begin{eqnarray}\label{INVFIELD}
 \widetilde A^{\mu}(x^0,{\bf x}^{\prime}) &=&  T^{\mu}_{ \;\alpha}(k^0,{\bf k} ) A^{\alpha}(x^0,{\bf x}^{\prime}).
\end{eqnarray}
Replacing (\ref{INVFIELD}) in (\ref{PARITYTRANS}) we obtain
\begin{eqnarray}
 \mathcal P\, \widetilde A^{\mu}(x^0,{\bf x}) \, \mathcal P^{-1}&=&
 \widetilde T^{\mu}_{ \;\alpha}(k^0,{\bf k} ) \widetilde A^{\alpha}(x^0,{\bf x}^{\prime}),
\end{eqnarray}
where the matrix of the transformation $ \widetilde T(k^0,{\bf k} )= R(k^0,{\bf k} ) T^ {-1}(k^0,{\bf k} )$ is find to be
\begin{equation} \label{NEWMATRIX}
 \left[
\begin{array}{cccccc}
1 & igk_2 &-igk_1  \\
-igk_2 & -1 &-igk_0 \\
 -igk_1& -igk_0 & 1
\end{array}%
\right]  \longrightarrow \left[
\begin{array}{cccccc}
1 & -g\partial^{\prime}_2 &g\partial^{\prime}_1  \\
g\partial^{\prime}_2 & -1 &g\partial^{\prime}_0 \\
 g\partial^{\prime}_1& g\partial^{\prime}_0 & 1
\end{array}%
\right].
\end{equation}
From (\ref{EFFECTIVEL}) and (\ref{NEWMATRIX}) we have that 
the Lagrangian transforms as
\begin{eqnarray}
\mathcal P \widetilde {\mathcal L}(x) \mathcal P^{-1} 
=\widetilde {\mathcal L}(x^{\prime})+\mathcal O(g^2),
\end{eqnarray}
where we have used $\widetilde T^T\, \eta \,\widetilde  T=\eta +\mathcal O(g^2)$. We conclude that the effective theory is invariant under parity transformations to the order considered.
One can in addition check that the effective Lagrangian is time reversal $\mathcal T$ and charge conjugation $\mathcal C$
invariant.
\section{Conclusions}
In this paper we have studied higher derivative field theories within a framework in which their corresponding higher 
derivative operators are regarded as small corrections to an ordinary-derivative field theory.
Historically, the use of higher derivative operators have been avoided in field theories
due to the many problems encountered with their formulation. 
Perhaps the most significant ones are the proliferation of extra degrees of freedom with respect to the ordinary ones
therefore going against the premise
that corrections should introduce small deviations together with  
the appearance of Hamiltonians being unbounded from below. Indeed, these problems arise regardless of whether
these corrections appear as free or interacting terms in the Lagrangian.
In general the solutions in the higher derivative theory can 
be classified according to their analytical behavior in the limit $g\to 0$.
We associate the analytical solutions to low-energy degrees of freedom and the nonanalytical solutions
to high-energy degrees of freedom.
In this regard, we see that the appropriate way of treating such corrections
as small perturbations is to implement techniques in order to control
and subsequently remove the high-energy degrees of freedom.
The perturbative formulation \cite{PERT-METHOD,STRING} 
accomplishes this by suppressing the excitation of high-energy modes
in a way consistent with the exact evolution and only allowing further
excitations of the low-energy modes already present in the zeroth order theory.

The main goal in this work has been to test the symmetries in the low-energy regime of higher derivatives field theories. 
In particular we have focused on the higher derivative Chern-Simons theory.
It is well known that symmetries in an effective theory will depend on the scheme
of approximation used to obtain the low-energy limit.
For example, the removal of all nonanalytical terms in the Hamiltonian (\ref{HAM2})
justified within an analytical expansion
would lead to the violations of Lorentz invariance \cite{HAMILTONIAN}. Also, we would have parity invariance to all orders with no
$g$ dependence by applying the equations of motion directly on
the extended MCS Lagrangian \cite{LORENTZ-INCONSISTENCIES}, which follows by replacing
the lowest order iterated solution $\Box A_{\alpha} \approx 0$
in the original Lagrangian.

To summarize, we have developed the Hamiltonian formulation
for the higher derivative MCS theory in $2+1$ dimensions. Given the singular nature
of the system we have been required to follow the Dirac method to impose second class constraints strongly from where we have
computed both the Dirac brackets and the reduced Hamiltonian.
In addition, we have exhibited in the Hamiltonian the negative-energies 
producing instabilities in the interacting case.
The perturbative method was implemented for the higher derivative Chern-Simons 
Lagrangian and an effective field theory that describes corrections to the low-energy physics with
a Hamiltonian well bounded from below and low-energy solutions were constructed.
We have introduced a prescription for the choice of the new canonical field variables which are unique according to the following
criteria: (i) the new gauge field transform
covariantly under Lorentz transformations and (ii) it satisfies the same gauge fixing condition as the old gauge field.
As suggested above, prior to (i) and (ii) there is some arbitrariness in the choice
of new canonical variables since the diagonalization of the symplectic form may be achieved 
by different sets of canonical variables.
Nevertheless, since they are connected by canonical
transformations they all describe physically equivalent theories.
Although it is beyond the scope of this paper, by
adopting similar arguments than before we can provide an expression
for the new gauge field to the next order 
which is uniquely given by
$ \widetilde A ^{\mu}= A^{\mu}+\theta g \epsilon^{\mu \alpha \beta}
\partial_ {\alpha }A_{\beta} + \kappa g^2\Box A^{\mu}  $ with $\kappa,\theta $ some real
numbers. Finally, we have obtained an effective field theory which is parity invariant to the order considered
therefore recovering symmetries one could expect at low energies.\appendix
\section*{Acknowledgments}
The author wants to thank L. F. Urrutia, J. D.
Vergara, H. A. M. Tecotl, D. Supanitsky and M. Cambiaso
for encouragement and useful comments
on this work. This research was supported by DGAPA-UNAM.
\section{Dirac Brackets}\label{APPENDIXA}
In this appendix we follow Dirac procedure to reduce second class
constraints from the higher derivative theory. To begin, let us
introduce the notation $\varphi_{A}(\chi_0,\chi_1 ,\varphi^i )$, with $A=\bar 0,\bar 1,1,2$.
The matrix of the second class constraints will be denoted by
\begin{eqnarray}
C_{AB}(t;{\bf x},{\bf x}^{\prime})=\{
\varphi_{A}(t,{\bf x}),\varphi_{B}(t,{\bf x}^{\prime}) \}.
\end{eqnarray}
From (\ref{CONSTRAINTALG}) we have
\begin{equation}
C_{AB}=\left[
\begin{array}{cccc}
0 & -1 & 0 & 0  \\
1 & 0 & -g\partial_2 &  g\partial_1 \\
0& -g\partial_2 & 0 & g  \\
0& g\partial_1 & -g & 0
\end{array}
\right] \delta ^{3}(\mathbf{x}-\mathbf{x}^{\prime}).
\end{equation}
The inverse matrix is
\begin{equation}
C_{AB}^{-1}=\left[
\begin{array}{cccc}
0 & 1 & -\partial_1 & -\partial_2 \\
-1 & 0 & 0 &  0 \\
-\partial_1& 0 & 0 & -1/g  \\
-\partial_2& 0 & 1/g & 0
\end{array}
\right] \delta ^{3}(\mathbf{x}-\mathbf{x}^{\prime}).
\end{equation}
The nonzero components are
\begin{align}
 C_{\bar 0\bar 1}^{-1}({\bf x},{\bf x}^{\prime})
 &=\delta ^{3}(\mathbf{x}-\mathbf{x}^{\prime}),\nonumber
\\
 C_{\bar 0i}^{-1}({\bf x},{\bf x}^{\prime})
 &=-\partial_i\delta ^{3}(\mathbf{x}-\mathbf{x}^{\prime}),
\\
 C_{ij}^{-1}({\bf x},{\bf x}^{\prime})
 &=-\frac{1}{g}\, \epsilon^{ij}\delta ^{3}
 (\mathbf{x}-\mathbf{x}^{\prime}),    \qquad i,j=1,2.\nonumber
\end{align}
And as usual Dirac brackets are defined as
\begin{equation}
\{ X, Y\}_D=\{ X, Y\}-\{ X, \varphi_A\}\;
C^{-1}_{AB}\;
 \{\varphi_B, Y\}.
\end{equation}
After some calculation the nonzero Dirac brackets are
\begin{align}
\{A_0( {t,\mathbf{x}}), \dot A_0(t,{\mathbf{x}}^{\prime} )\}_
D&=-\delta(\mathbf{x}- \mathbf{x}^{\prime}),\nonumber
\\
\{ A_0({t,\mathbf{x}}), P_0( {t,\mathbf{x}^{\prime}} )\}_D&=\delta(\mathbf{x}- \mathbf{x'}),\nonumber
\\
\{ A_0({t,\mathbf{x}}), P^i(t,{\mathbf{x}}^{\prime})\}_D
&=\frac{g}{2}\epsilon^{ij}\partial_j\, \delta(\mathbf{x}-
 \mathbf{x}^{\prime}),\nonumber
\\
\{ \dot A_i( {t,\mathbf{x}}), \dot A_j(t,{\mathbf{x} }^{\prime})\}_D
&=-\frac{1}{g}\epsilon^{ij}\delta(\mathbf{x}- \mathbf{x}
^{\prime}),\nonumber
\\
\{\dot  A_i({t,\mathbf{x}}), \dot A_0 (t,{\mathbf{x}}
^{\prime})\}_D
&=-\partial_i\,\delta(\mathbf{x}- \mathbf{x}^{\prime}),\nonumber
\\
\{\dot  A_i({t,\mathbf{x}}), P_0 (t,{\mathbf{x}}^{\prime})\}_D
&=\frac{1}{2}\partial_i\,\delta(\mathbf{x}- \mathbf{x}^{\prime}),
\nonumber
\\
\{\dot  A_i({t,\mathbf{x}}), P^j (t,{\mathbf{x}}^{\prime})\}_D&
=\frac{g}{2}\epsilon^{jk}\partial_i\partial_k\delta(\mathbf{x}-
\mathbf{x}^{\prime}),\nonumber
\\
\{\dot  A_i({t,\mathbf{x}}), \Pi^j (t,{\mathbf{x}}^{\prime})\}_D
&=\frac{1}{2}\delta_{ij}\,\delta(\mathbf{x}- \mathbf{x}
^{\prime}),
\\
\{\dot  A_0({t,\mathbf{x}}), P^i (t,{\mathbf{x}}^{\prime})\}_
D&=\partial_i\,\delta(\mathbf{x}- \mathbf{x}^{\prime}),\nonumber
\\
\{P_0({t,\mathbf{x}}), \Pi^i (t,{\mathbf{x}}^{\prime})\}_D& \frac{g}{4}\epsilon^{ij}\partial_j\,\delta(\mathbf{x}-
 \mathbf{x}^{\prime}),\nonumber
\\
\{\Pi_0({t,\mathbf{x}}), P^i (t,{\mathbf{x}}^{\prime})\}_D&= -\frac{g}{2}
\epsilon^{ij}\partial_j\,\delta(\mathbf{x}- \mathbf{x}^{\prime}),
\nonumber
\\
\{P^i({t,\mathbf{x}}), P^j (t,{\mathbf{x}}^{\prime})\}_D&\frac{g}{2}\epsilon^{ij}  \nabla^2
\delta(\mathbf{x}- \mathbf{x^{\prime}}),\nonumber
\\
\{\Pi^i({t,\mathbf{x}}), \Pi^j (t,{\mathbf{x}}^{\prime})\}_D
&= -\frac{g}{4}\epsilon^{ij} \delta(\mathbf{x}- \mathbf{x}
^{\prime}),\nonumber
\end{align}
and the reduced Hamiltonian density is
\begin{align}
\mathcal H_R=&
P_0 \dot A_0 -\frac{2}{g} \Pi_i \left(\epsilon^ {ij}P_j
+\frac{\Pi_i}{g}+\frac{g}{2} \nabla^2A_i \right)\nonumber \\ &
-\partial_i A_0 \left(P_i+\frac{g}{2}\epsilon^{ij} \nabla^2A_j \right)
+\frac{1}{4}F_{ij}^2+\frac{1}{2}(\partial 
\cdot A )^2.\nonumber \\
\end{align}
Finally, the reader may check that the reduced Hamiltonian
together with the Dirac brackets give the exact higher
derivative equations of motion.


\begin{thebibliography}{99}
\bibitem{COMMENT1}
The same solutions (\ref{MODES}) are found when the extended MCS theory
is considered without the Lorentz gauge fixing term. However, in that case
the equation of motion (\ref{MAXWELL}) is changed by
$ \Box \left( 1 +g^2 \Box   \right)  F^{*\mu}=0$ where
$F^{*\mu}$ is the pseudovector dual field.
\bibitem{COMMENT4}
Considering the four second class
constraints and the six configuration variables we have $(2 \times 6-4)/2=4$ total degrees
of freedom. This exceeds in three the number of physical degrees of freedom
in the Maxwell theory in two spatial dimensions: one extra degree of freedom due to the massive
propagation and associated to the higher derivative term and two more due to the
time and longitudinal propagations appearing in the Lorentz gauge.
\end{thebibliography}
\end{document}